\documentclass[pra,aps,showpacs,nofootinbib,twocolumn,superscriptaddress]{revtex4}
\usepackage{amssymb}
\usepackage{graphicx}

\input{tcilatex}
\begin{document}

\title{Multiple-quantized vortices in rotating LOFF state of ultracold Fermi
superfluid gas}
\author{Miodrag L. Kuli\'{c}}
\affiliation{Institute for Theoretical Physics, Goethe-University D-60438 Frankfurt am
Main, Germany}
\author{Armen~Sedrakian}
\affiliation{Institute for Theoretical Physics, Goethe-University D-60438 Frankfurt am
Main, Germany}
\author{Dirk H. Rischke}
\affiliation{Institute for Theoretical Physics, Goethe-University D-60438 Frankfurt am
Main, Germany}
\affiliation{Frankfurt Institute for Advanced Studies, D-60438 Frankfurt am Main, Germany}

\begin{abstract}
A rotating ultracold $S$-wave superfluid Fermi gas is considered, when the
population imbalance $\delta n$ (or equivalently the mismatch in chemical
potentials $\delta \mu $) corresponds to the
Larkin-Ovchinnikov-Fulde-Ferrell (LOFF) state in the vicinity of the
Lifshitz critical point. It is shown that under these conditions the
critical angular velocity $\Omega _{c2}$ in two-dimensional systems is an
oscillating function of temperature and $\delta n$ giving rise to reentrant
superfluid phases. This leads to vortex lattices with multiple-quantized
circulation quanta. The reason for this behavior is the population by Cooper
pairs of the Landau levels above the lowest one.
\end{abstract}

\date{\today }
\maketitle

\section{Introduction}

The Larkin-Ovchinnikov-Fulde-Ferrel (LOFF) phase with a spatial variation of
the superconducting order parameter $\Delta (\mathbf{r})$ can be realized in
superconductors in external magnetic field ($h$). The Zeeman interaction
between spins of electrons and the external field $h$ produces unequal
populations of the spin up and down electrons. This, in turn implies a
mismatch in chemical potentials $\delta \mu =h\equiv (\mu _{\uparrow }- \mu
_{\downarrow })/2$ of spin-up and down components~\cite{LOFF,FFLO}.

In ultracold fermionic superfluids the pairing occurs between atoms in two
different hyperfine states ($a,b $), which can be loaded into a trap with
unequal numbers. The situation is then analogous to the spin-polarized
electrons in metallic superconductors. Due to the rather large relaxation
time for transitions between the different hyperfine states, the numbers of
both species $n_{a}$ and $n_{b}$ can be considered as fixed to a good
approximation. Since the particle losses are negligible, the population
imbalance $\delta n=n_{a}-n_{b}$ is fixed as well. The latter implies that
the chemical potentials are mismatched; this mismatch is commonly
characterized by the quantity $\delta\mu =(\mu _{a}-\mu _{b})/2$. The
possibility to fix $\delta n$ means that the physical realization of the
LOFF phase in a superfluid ultracold Fermi gas is more favorable than in
metallic superconductors.

In metallic $S$-wave superconductors placed in an external magnetic field
the genuine LOFF state is usually masked by the orbital effects due to the
Lorenz force term in the Hamiltonian described by the minimal coupling
prescription: $\mathbf{\hat{p}}\rightarrow \mathbf{\hat{p}}-e\mathbf{A}$,
where $\mathbf{\hat{p}}$ and $e$ are the electron momentum and charge, $%
\mathbf{A}$ is the vector potential. To eliminate the orbital effects one
needs specially designed materials like in a clean quasi-2D superconductor
which is exposed to a magnetic field applied parallel to the conduction
planes. Another example are the heterogeneous structures, where the LOFF
phase can be realized in a much broader range of parameters than in bulk
superconductors or superfluids. Specifically, in metallic
superconductor-ferromagnet-superconductor (\textit{SFS}) weak links the LOFF
phase has already been realized within the ferromagnetic link. These systems
are very promising for small-scale applications, such as the \textit{SQUID}%
s, quantum computing, etc.

The complications mentioned above does not arise in superfluid two-component
ultracold Fermi gases, where the number density is controlled in
experiments, and orbital effects are absent due to the neutrality of gases.
However, the realization of the LOFF phase in superfluid ultracold Fermi
gases is a still unresolved issue. These systems are commonly placed in
magnetic traps that create an inhomogeneous confining potential which plays
the role of the container that keeps the ultracold gas in equilibrium. In
such an inhomogeneous system a nonzero population imbalance leads to a phase
separation into a homogeneous superfluid (\textit{S}) and normal (\textit{N}%
) fluid. This leaves a rather narrow spatial window between \textit{S} and
\textit{N} phases where the LOFF phase can be realized~\cite{Machida-Torma}.
The optical lattices may offer more favorable conditions for the realization
and applications of the LOFF phase in ultracold gases~\cite{Koponen}. One
possible application, proposed in Ref.~\cite{Kulic}, is the construction of
a heterogeneous \textit{SNS} weak link, where \textit{N} is an ultracold
Fermi gas in the normal state with population imbalance $\delta n_N\neq 0$
(and $\delta \mu _{N}\neq 0$), while the superfluid banks \textit{S} are
from superfluid ultracold gas (of the same atoms), but without population
imbalance, i.e., $\delta n_{S}=0$. It turns out that in such a case,
depending on parameters of the system, one can in principle realize a $\pi$%
-weak link (a type of Josephson contact with the phase difference $\varphi
=\pi $ in the ground state), which when placed in a superfluid ring can
produce spontaneous mass flow in the ground state - the so called $\pi
-SQUID $~\cite{Kulic}. Since the LOFF state is characterized by a negative
stiffness (see below) which causes periodically modulated superconducting
(superfluid) order parameter $\Delta (\mathbf{r}+\mathbf{L})=\Delta (\mathbf{%
r})$, this fact may gives rise to a number of interesting effects.

In this work we consider 2D systems with the LOFF phase in a two-component
ultracold Fermi gas under rotation. The angular rotation of superfluid gases
plays a role that is similar to the orbital effect in metallic
superconductors. This fact opens up the possibility to investigate in
ultracold superfluid Fermi gases the interplay of \textquotedblleft orbital"
and \textquotedblleft spin" effects. In contrast to metallic superconductors
these quantities can be varied in ultracold gases independently. Below we
investigate the behavior of the critical angular velocity $\Omega _{c2}$ for
the transition from the superfluid to the normal state in the vicinity of
the Lifshitz critical point (with the temperature $T^{\ast }$ and the
chemical potential mismatch $\delta \mu ^{\ast }$ in the $T-\delta \mu $
plane). We show that $\Omega _{c2}$ is an oscillatory function of
temperature or population imbalance. These oscillations arise due to the
occupation of (higher) Landau levels by Cooper pairs. The latter effect is
realized if the quasiclassical condition $\hbar \Omega _{c2}<\pi
(k_{B}T_{c})^{2}/\mu $ is realized, where $T_{c}$ is the critical
temperature of superfluid phase transition and $\mu $ is the average
chemical potential. We refer to this effect as the \textit{quasiclassical
oscillation effect}. In the opposite case, more precisely when $\hbar \Omega
_{c2}>T_{c}$, the quantity $\Omega _{c2}$ oscillates due to the population
of the Landau levels in the normal state (the normal state Landau
quantization)~\cite{Gruenberg}, \cite{Bulaevskii}, \cite{Sedrakian}. This
effect, which can be refered to as the \textit{quantum oscillation effect}
is negligible near the point $(T^{\ast },\delta \mu ^{\ast })$ and will not
be studied here.

\section{Rotating the LOFF phase}

We consider a rotating Fermi gas with the angular velocity vector directed
along the z-axis, $\mathbf{\Omega }=\Omega \mathbf{\hat{z}}$. The ultracold
gas is placed in a magnetic trap with the potential $V_{m}(\mathbf{r}%
)=[M\omega ^{2}\mathbf{r}^{2}+M\omega _{z}^{2}z^{2}]/2$, where $\omega$ and $%
\omega_z$ are the trapping frequencies, $M$ is the atomic mass. The BCS
Hamiltonian \textit{in the rotating coordinate system}, but \textit{%
expressed via coordinates of the inertial laboratory system}, is given by
\[
\hat{H}=\sum_{i=a,b}\int d^{d}x\hat{\psi}_{i}^{\dag } [h_{0}(\mathbf{\hat{p}}%
, \mathbf{r},\mathbf{\hat{L}})-\mu _{i}]\hat{\psi}_{i}
\]
\begin{equation}
-g\int d^{d}x\hat{\psi}_{a}^{\dag }(x)\hat{\psi}_{b}^{\dag }(x)\hat{\psi}%
_{b}(x)\hat{\psi}_{a}(x),  \label{H}
\end{equation}
where the summation is over the fermionic species, the integration involves
the dimension of the space $d$, $\hat{\psi}^{\dag}_{a}(x)$ and $\hat{\psi}%
_{a}(x)$ are the fermionic creation and annihilation operators, $g$ is the
coupling constant, and $h_0$ is the single particle Hamiltonian, which will
be specified below. The chemical potentials of the species are $\mu
_{a,b}=\mu \pm \delta \mu $, where $\mu =(\mu _{a}+\mu _{a})/2$ and $\delta
\mu =(\mu _{a}-\mu _{b})/2$ are the average and the ``mismatch'' chemical
potentials, respectively. In the following we specify the quantities $\mu $
and $\delta\mu$. Since the self-consistent equation for the
superconducting(superfluid) order parameter are the same for fixed $%
\delta\mu $ and $\delta n$ the obtained results are applicable also to
systems with fixed $\delta n$. The single-particle Hamiltonian $h_{0}(%
\mathbf{\hat{p}},\mathbf{r},\mathbf{\hat{L}})$ in Eq. (\ref{H}) is given by
\begin{equation}
h_{0}(\mathbf{\hat{p}},\mathbf{r},\hat{L}_{z})=\frac{\mathbf{\hat{p}}^{2}}{2M%
}+V_{m}(\mathbf{r})-\mathbf{\Omega }\cdot \mathbf{\hat{L}},  \label{h0}
\end{equation}%
where $\mathbf{\hat{p}}=-i\hbar \nabla $ is the linear and $\mathbf{\hat{L}}=%
\mathbf{\hat{r}}\times \mathbf{\hat{p}}$ is the orbital momentum. If
necessary $h_{0}(\mathbf{\hat{p}},\mathbf{r},\mathbf{\hat{L}})$ comprises
also the optical-lattice potential $V_{op}(\mathbf{r})$, i.e. one has
\[
h_{0}(\mathbf{\hat{p}},\mathbf{r},\hat{L}_{z})=\frac{1}{2M}(\mathbf{\hat{p}}%
-M\mathbf{V}_{\Omega })^{2}
\]%
\begin{equation}
+\frac{1}{2}M(\omega ^{2}-\Omega ^{2})r^{2}+\frac{1}{2}M\omega _{z}^{2}z^{2},
\label{h0-V}
\end{equation}%
where $\mathbf{V}_{\Omega }=\Omega \mathbf{\hat{z}\times r}$ is the velocity
due to the rotation. In the following we consider the \textit{%
two-dimensional (2D) gas} which is realized for the pancake-like trap with $%
\omega _{z}\gg \omega $. It is also assumed that $\omega \gtrsim \Omega $ in
order to keep the gas stable and quasi-homogeneous as much as possible. When
necessary, the system can be additionally placed in a periodic
optical-lattice potential $V_{op}(\mathbf{r})$, which changes the
atomic-particle spectrum from the parabolic one $\mathbf{p}^{2}/2$ to the
tight-binding like $\epsilon (\mathbf{\hat{p}})$, which creates favorable
conditions for the LOFF phase \cite{Koponen}. We study the problem near the
second-order transition line between the normal (unpaired) and the LOFF
phase in the $T-\delta \mu $ plane, which is realized for $T\leqslant
T^{\ast }$ and $\delta \mu \geqslant \delta \mu ^{\ast }$ \cite{Rainer}. It
will be shown that near this line the critical rotation velocity $\Omega
_{c2}$ (for the transition from the normal to the superfluid state) is small
and the effect of rotation can be accounted for in the quasiclassical
approximation. The latter means that the Landau quantization in the normal
Fermi gas is negligible in this parameter region. At this line the equation
for the order parameter can be linearized
\begin{equation}
\frac{\Delta (\mathbf{r})}{g}=\int d^{2}r_{1}K_{0}(\mathbf{r}-\mathbf{r}%
_{1})e^{-i(\mathbf{r}-\mathbf{r}_{1})(\mathbf{\hat{p}}-2M\mathbf{V}_{\Omega
})}\Delta (\mathbf{r})  \label{Delta}
\end{equation}
with the kernel

\begin{equation}
K_{0}(\mathbf{r}-\mathbf{r}_{1})=T\sum_{\omega _{n}}G_{a}(\omega _{n},%
\mathbf{r-r}_{1})G_{b}(-\omega _{n},\mathbf{r-r}_{1}),  \label{K-0r}
\end{equation}%
where $G_{a/b}(\omega _{n},\mathbf{r-r}_{1})$ are the single particle
Green's functions, where $\hbar \omega _{m}\equiv \eta _{m}=\pi k_{B}T(2m+1)$
are the fermionic Matsubara frequencies. The characteristic scale of the
kernel $K_{0}(\mathbf{r}-\mathbf{r}_{1})$ is given by the superfluid
coherence length $\xi _{0}$. Therefore, the quasiclassical approximation in
Eq.~(\ref{Delta}) is valid if the phase change $\delta \varphi $ due to the
rotation is small on the scales of the order of $\xi _{0}$. The phase change
can be estimated as
\begin{equation}
\delta \varphi =2M\int_{\mathbf{r}}^{\mathbf{r+\xi }_{0}}\mathbf{V}_{\Omega
}(\mathbf{l})d\mathbf{l}\sim \frac{2\hbar \Omega \mu }{(k_{B}T_{c})^{2}}.
\label{delta-phi}
\end{equation}%
Then, the quasiclassical condition $\delta \varphi \ll 2\pi $ gives $\mathbf{%
\hbar }\Omega \ll \pi (k_{B}T_{c})^{2}/\mu $ where $\mu =E_{F}$ and $E_{F}$
is the (average) Fermi energy. When $\Omega $ is large enough so that the
cyclotron radius $R_{\Omega }\sim v_{F}/\Omega $ of the atomic orbits is
much smaller than $\xi _{0}$, i.e., when $\mathbf{\hbar }\Omega \gtrsim
k_{B}T_{c}$, it is necessary to take into account the Landau quantization of
atomic motion in the normal state.

Consider next the Fourier-image of the kernel
\begin{equation}
K_{0}(\mathbf{q})=k_{B}T\sum_{m}\int d^{2}pG_{a}(\omega _{m},\mathbf{p}%
)G_{b}(-\omega _{m},\mathbf{p}+\mathbf{q}).  \label{K-0q}
\end{equation}%
In the quasi-homogeneous case, when the effect of the potential $M(\omega
^{2}-\Omega ^{2})r^{2}/2$ on the single-particle spectrum is small, the
Green's functions $G_{a,b}$ are given by
\begin{equation}
G_{a,b}(\omega _{m},\mathbf{p})=(i\eta _{m}-\xi (\mathbf{p})\pm \delta \mu
)^{-1},  \label{Gab-q}
\end{equation}%
where $\xi (\mathbf{p})=(\mathbf{p}^{2}/2M)-\mu $. After the integration
over the energy $\xi $ and by assuming a circular Fermi surface one obtains
(in the following $k_{B}=1$)
\begin{equation}
K_{0}(\mathbf{q},\eta _{m}>0)=iN(0)\int_{0}^{2\pi }\frac{d\varphi }{2(i\eta
_{m}+\delta \mu )+qv_{F}\cos \varphi },  \label{Kq-fi}
\end{equation}%
where $v_{F}$ is the Fermi velocity and $N(0)$ is the density of states at
the Fermi surface. The integration in Eq.~(\ref{Kq-fi}) is straightforward
(the contour integral runs over the contour $|z|=1$ with $z=\exp \{i\varphi
\}$) and we obtain
\begin{equation}
K_{0}(q)=\mathrm{\func{Re}}\left( \sum_{m}\frac{\pi N(0)T}{\sqrt{(\eta
_{m}+i\delta \mu )^{2}+(\hbar qv_{F}/2)^{2}}}\right) .  \label{K0q-expl}
\end{equation}%
Since we consider the problem near the Lifshitz critical point $(T^{\ast
},\delta \mu ^{\ast })$, the wave vector of the LOFF phase is small, i.e., $%
q\ll \xi _{0}^{-1}$, and it is sufficient to make the small $q$ expansion.
Thus, we approximate $K_{0}(\mathbf{q})\approx K_{0}(0)+K_{2}q^{2}+K_{4}q^{4}
$, with $K_{0}(0)=N(0)[g^{-1}-\tau (t,\delta \bar{\mu})]$, where $\tau
(t,\delta \bar{\mu})$ is defined in Eq.(\ref{tau}). Upon defining
dimensionalless unit $X=(q^{2}2M)/E_{c}$ with $E_{c}=T_{c}^{2}/E_{F}$, we
rewrite this expansion as $K_{0}(\mathbf{q})\approx
K_{0}(0)+k_{2}X+k_{4}X^{2}$, where
\begin{eqnarray}
k_{2} &=&\frac{t^{-2}}{(4\pi )^{2}}\mathrm{\func{Re}}\psi ^{(2)}\left( \frac{%
1}{2}+i\frac{\delta \bar{\mu}}{2\pi t}\right) ,  \label{k2} \\
k_{4} &=&-\frac{t^{-4}}{4(4\pi )^{4}}\mathrm{\func{Re}}\psi ^{(4)}\left(
\frac{1}{2}+i\frac{\delta \bar{\mu}}{2\pi t}\right) ,  \label{k4}
\end{eqnarray}%
and%
\begin{equation}
\tau (t,\delta \bar{\mu})=\ln t+\mathrm{\func{Re}}\psi \left( \frac{1}{2}+i%
\frac{\delta \bar{\mu}}{2\pi t}\right) -\psi \left( \frac{1}{2}\right) .
\label{tau}
\end{equation}%
Here, $t=(T/T_{c0})$ and $\delta \bar{\mu}=\delta \mu /T_{c0}$, where $T_{c0}
$ is the critical temperature of superfluid phase transition when the number
densities of the two species are equal ($\delta n=0$), and $\psi (x)=d\ln
\Gamma (x)/dx$ is di-gamma function and $\psi ^{(n)}(x)=d^{n}\psi (x)/dx^{n}$%
. In case of $\Omega =0$ one obtains from Eq.~(\ref{Delta}) the equation
defining the transition line between the normal and the LOFF phase
\begin{equation}
\lbrack \tau _{L}(t_{Q},\delta \bar{\mu})-k_{2}E(Q)-k_{4}E^{2}(Q)]=0,
\label{LOFF}
\end{equation}%
where $\Delta (r)=\Delta _{q}\exp [\mathbf{Q}\cdot \mathbf{r}]\neq 0$ and $%
E(Q)\equiv (2Q^{2}M)/E_{c}=k_{2}/2|k_{4}|$. The magnitude of the LOFF wave
vector $Q$ is determined by the maximum of the function $%
k_{2}E(Q)+k_{4}E^{2}(Q)$, i.e., by maximizing the LOFF critical temperature $%
t_{L}(Q,\delta \bar{\mu})$. The direction of the wave-vector is chosen
spontaneously. At the Lifshitz point where $t^{\ast }=0.56$ and $\delta \bar{%
\mu}^{\ast }=1.04$ one obtains $\tau _{L}(t^{\ast },\delta \bar{\mu}^{\ast
})=0$ and $k_{2}(t^{\ast },\delta \bar{\mu}^{\ast })=0$, where $Q(t^{\ast
},\delta \mu ^{\ast })=0$. The LOFF phase is realized for $t<t^{\ast }$and $%
\delta \bar{\mu}>\delta \mu ^{\ast }$, therefore near the Lifshits point one
has $Q(t,\delta \bar{\mu})\ll \xi _{0}^{-1}$, which justifies the small-$q$
expansion of the kernel $K_{0}(\mathbf{q})$ above.

In the case of a 2D rotating system, with $\Omega \neq 0$and $\mathbf{V}%
_{\Omega }=\mathbf{\Omega }\times \mathbf{r}$, there is an upper critical
angular velocity $\Omega _{c2}(t,\delta \bar{\mu})$ below which there is a
nucleation of superfluidity in the form of quantized vortices. The linear
Ginzburg-Landau equation for the \textit{second-order normal state - LOFF
transition} reads
\begin{equation}
\left[ \tau (t,\delta \bar{\mu})-\left( \frac{k_{2}}{E_{c}}\right) \hat{H}%
-\left( \frac{k_{4}}{E_{c}^{2}}\right) \hat{H}^{2}\right] \Delta (\mathbf{r}%
)=0,  \label{Omega-c}
\end{equation}%
where $\hat{H}\equiv \left( \mathbf{\hat{p}}-2M\mathbf{V}_{\Omega }\right)
^{2}/2M$ is the Hamiltonian for the harmonic oscillator in the isotropic
Coulomb gauge. To obtain $\Omega _{c2}(t,\delta \bar{\mu})$ we need the
solution of the eigenvalue problem $\hat{H}\Delta _{n}(\mathbf{r})=\epsilon
_{n}\Delta _{n}(\mathbf{r})$. By assuming a rotationally infinite 2D system,
such as disc, the solution of the eigenvalue problem is given by $\Delta
_{n}(\mathbf{r})\sim z^{n}\exp \{-(|z|/2l_{0})^{2}\}$, with $z=x+iy$ and the
\textquotedblleft magnetic" length $l_{0}=\sqrt{\hbar /4M\Omega _{c2}}$,
while the eigenvalues are $\epsilon _{n}=4\Omega _{c2}(n+1/2)$, $n=0,1,2..$.
Note that the solutions $\Delta _{n}(\mathbf{r})$ are also eigenstates of
the $z$-the component of the angular momentum operator $\hat{L}_{z}=-i\hbar
(x\partial _{y}-y\partial _{x})\equiv \hbar (z\partial _{z}-z^{\ast
}\partial _{z^{\ast }})$, i.e.%
\begin{equation}
\hat{L}_{z}\Delta _{n}(\mathbf{r})=\hbar n\Delta _{n}(\mathbf{r}).
\label{Lz}
\end{equation}%
The solutions with $n=1,2...$ may correspond to a multiply-quantized vortex
lattice while the case with $n=0$ corresponds to the standard Abrikosov
solution where the number of the zeros of the order parameter and flux
quanta in the unit cell is one. However, the calculation of the real
structure of the vortex lattice for $\Omega <\Omega _{c2}$ requires the
knowledge of the nonlinear Ginzburg-Landau equation and the superconducting
current in the LOFF state, which are much more complicated than the standard
equations in the homogeneous case \cite{Buzdin-Kulic}. We expect that in the
rotating LOFF state with $n=1,2...$ various vortex lattices may appear,
where the number of zeros of $\Delta (\mathbf{r})$ in the unit cell can be
larger than the number of flux quanta~\cite{Klein}. The solution of the
eigenvalue problem posed by Eq.~(\ref{Omega-c}) gives the explicit
expression $\Omega _{c2,\pm }^{(n)}$ for given $n$
\begin{equation}
\frac{\Omega _{c2,\pm }^{(n)}(t,\delta \bar{\mu})}{\omega _{c}}=\frac{1}{n+%
\frac{1}{2}}\frac{k_{2}\pm \sqrt{k_{2}^{2}-4\tau |k_{4}|}}{8|k_{4}|}.
\label{Omega-c2}
\end{equation}%
where $\omega _{c}=E_{c}/\hbar $. It is seen that for the \textit{fixed value%
} of $\delta \bar{\mu}>\delta \bar{\mu}^{\ast }$ (or for fixed $\delta n$)
one has two branches $\Omega _{c2,+}^{(n)}>\Omega _{c2,-}^{(n)}$ which are
functions of the temperature $t<t^{\ast }$. They meet each other on the line
$k_{2}^{2}(t,\delta \bar{\mu})-4\tau (t,\delta \bar{\mu})|k_{4}(t,\delta
\bar{\mu})|=0$, which is in fact the LOFF line $t_{L}(t_{Q},\delta \bar{\mu})
$ given by Eq.~(\ref{LOFF}). The lower line for the $n$-th level intersects
the upper line for the $n+1$-th level at the point $\Omega
_{c2,-}^{(n)}(t_{n,n+1})=\Omega _{c2,+}^{(n+1)}(t_{n,n+1})$. This means that
in the temperature interval $t_{n,n+1}<t<t_{L}$ \ (for fixed $\delta \bar{\mu%
}$) the \textit{normal state is realized} for $\Omega
_{c2,+}^{(n+1)}(t)<\Omega <\Omega _{c2,-}^{(n)}(t)$. Therefore, a cascade of
oscillatory (interchanging) normal and LOFF-vortex-lattice \textit{reentrant
transitions} will arise, i.e. a sequence of transition from the normal
ultracold Fermi gas to the superfluid gas with multiply-quantized vortex
lattice. This is clearly seen in the phase diagram $\Omega _{c2}(t)$ shown
in Fig. 1.

\begin{figure}[tbp]
\resizebox{.4\textwidth}{!} {\includegraphics*
[width=6cm]{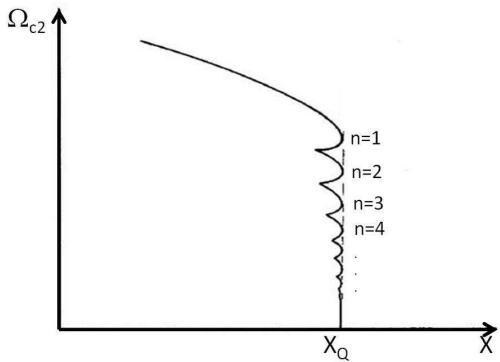}} \caption{Schematic figure of the
oscillation of the critical angular velocity $\Omega _{c2}(X;y)$
as the function of the \textit{variable} $X$ for \textit{fixed}
$y$. For the variable temperature $t=T/T_{c0}$ and fixed mismatch
in chemical potentials $\protect\delta \bar{\protect\mu}$ (or the
population imbalance $\protect\delta n$) one has $X=t$ and
$y=\protect\delta \bar{\protect\mu}$ (or $y=\protect\delta n$) and
vice versa. $X_{Q}$ is
the LOFF critical temperature $t_{L}$ or the critical mismatch $\protect%
\delta \bar{\protect\mu _{L}}$.}
\end{figure}

We would like to stress that the phase diagram in Fig. 1 is also generic for
the case when the temperature $t<t^{\ast }$ is fixed but the mismatch in
chemical potentials $\delta \bar{\mu}>\delta \bar{\mu}^{\ast }$ (or $\delta n
$) is varied. In that case the variables $t$ and $\delta \bar{\mu}$ change
the roles but the oscillatory effect is the same. Again one has $\Omega
_{c2,+}^{(n)}(\delta \bar{\mu})>\Omega _{c2,-}^{(n)}(\delta \bar{\mu})$ and
these two curves meet at the LOFF transition line $\delta \bar{\mu}_{Q}(t)$.
In the intervals $\Omega _{c2,+}^{(n+1)}(\delta \bar{\mu})<\Omega <\Omega
_{c2,-}^{(n)}(\delta \bar{\mu})$ one has again the cascade of normal to the
LOFF-vortex-lattice reentrant transitions.

\section{Discussion and conclusions}

Our study shows that it is possible to create a rotating LOFF state in
superfluid ultracold Fermi gases in a quasi-2D magnetic trap, if the
trap-frequency $\omega $ is adapted to be around (but still larger than) the
angular velocity $\Omega _{c2}(T,\delta \mu _{r})$ given by Eq.~(\ref%
{Omega-c2}), i.e., $\omega >\Omega _{c2}$. The latter condition allows,
first, the stability of the ultracold gas and second it gives rather small
inhomogeneity in the quasiparticle spectrum. This means that the kernel $%
K_{0}(\mathbf{r})$ in Eq.~(\ref{K-0r}) can be calculated with the help of
the Green's functions for the homogeneous system given by Eq.~(\ref{Gab-q}).
The realization of the LOFF phase is even more favorable if the ultracold
gas is placed in a 2D optical lattice, since the existence of the van Hove
singularities near (or at) the Fermi surface can favor the LOFF state in a
much broader region of the phase diagram $(\delta n,T)$~\cite{Koponen}. In
the case of the rotating LOFF superfluid an additional weak magnetic trap,
with $\omega \gtrsim \Omega _{c2}$, compensates again the centrifugal
potential due to the rotation.

The order of magnitude of $\Omega _{c2}^{(n)}$ near the Lifshitz point is $%
\Omega _{c2}^{(n)}\sim (10^{-1}-10^{-2})(T_{c}^{2}/E_{F})$ and in the case
of the superfluid Li$^{6}$ where $E_{F}\sim 2$ $\mu K$ and $T_{c}\sim
E_{F}/10$ one has $\Omega _{c2}^{(n)}<(30-300)$ $s^{-1}$. These values are
in the range of the experimentally reached rotational frequencies realized
by stirring methods for creating vortices~\cite{Madison}. Finally, we stress
that the whole analysis was carried out in the framework of the BCS weak
coupling theory. We anticipate that our analysis can be extended towards the
Bose-Einstein condensate limit, and it should remain (qualitatively) valid
at stronger couplings, but on the BCS side of the phase diagram. This
expectation is justified by the studies of the vortex state in metallic
superconductors, for which the fermionic excitations in the vortex cores are
realized also in the strong coupling limit but still on the BCS side of the
phase diagram.

In conclusion, we have studied a rotating two-component superfluid ultracold
Fermi gas with a mismatch in chemical potentials of the species (or with
population imbalance), which is in the LOFF phase and near the Lifshitz
point. We have shown that the critical angular velocity $\Omega _{c2}$ in
two-dimensional systems is an oscillatory function of temperature or the
population imbalance $\delta n(=n_{a}-n_{b})$. This effect gives rise to a
cascade of reentrant superfluid transitions with the superfluid featuring
multiple-quantized circulation quanta. The reason for this oscillatory
effect is the population of the higher Landau levels ($n\geqslant 1$) by
Cooper pairs in the LOFF state. The obtained results on the critical angular
velocity $\Omega _{c2}$ in the LOFF state might be of interest also for
rotating 2D metallic superconductors in the parallel magnetic field and for
color superconductors in quark matter, where in the two-flavor
color-superconducting quark matter the LOFF phase may compete with the
gluonic condensate~\cite{Rischke}.

\textbf{Acknowledgement}. M. L. K. acknowledges the support by the Deutsche
Forschungsgemeinschaft (Grant $SE$ $1836/1-1$).


\begin{references}

\bibitem{LOFF}
A. I. Larkin, Yu. N. Ovchinikov,
Zh. Eksp. Teor. Fiz. {\bf 47}, 1136 (1964)
[Sov. Phys. JETP {\bf 20}, 762 (1965)]

\bibitem{FFLO}
P. Fulde, R. Ferrell,
Phys. Rev.  {\bf 135}, A550 (1964)

\bibitem{Machida-Torma}
J. Kinnunen et al., Phys. Rev. Lett. {\bf 96}, 110403(2006); K.
Machida et al., Phys. Rev. Lett. {\bf 97}, 120407(2006)

\bibitem{Koponen}
T. K. Koponen et al., Phys. Rev. Lett. {\bf 99}, 120403(2007)

\bibitem{Kulic}
M. L. Kuli\'{c}, Phys. Rev.  {\bf A 76}, 053625 (2007)

\bibitem{Gruenberg}
L. W. Gruenberg, L. Gunther, Phys. Rev. Lett. {\bf 16}, 996(1966);
L. Gunther, L. W. Gruenberg, Solid St. Comm. {\bf 4}, 329(1966)

\bibitem{Bulaevskii}
L. N. Bulaevskii, Sov. Phys. JETP {\bf 37}, 1133 (1973)

\bibitem{Sedrakian}
 A. Sedrakian, J. Mur-Petit, A. Polls, H. M\"uther,
Phys. Rev. A {\bf 72} (2005) 013613.


\bibitem{Rainer}
H. Burkhard, D. Rainer, Ann. Phys. (Leipzig) {\bf 3}, 181(1994)

\bibitem{Buzdin-Kulic}
A. I. Buzdin, M. L. Kuli\'{c}, J. Low Temp. Phys. {\bf 54}, 203
(1984); M. L. Kuli\'{c}, U. Hofmann, Solid St. Comm. {\bf 77},
717(1991)

\bibitem{Klein}
U. Klein, H. Shimahara, D. Rainer, J. Low Temp. Phys. {\bf 118}, 91
(2000); M. Houzet, A. I. Buzdin, Europhys. Lett.,{\bf 50} (3), p.375 (2000)

\bibitem{Madison}
K. W. Madison et al., Phys. Rev. Lett. {\bf 84}, 806 (2000)

\bibitem{Rischke}
O. Kiriyama, D. H. Rischke, I. A. Shovkovoy,
Phys. Lett. {\bf B 43}, 331 (2006)

\end{references}
\end{document}